\documentclass[twocolumn,aps,prl,reprint,groupedaddress]{revtex4}
\usepackage{graphicx}
\usepackage{color}

\newcommand{\be}{\begin{equation}}
\newcommand{\ee}{\end{equation}}
\newcommand{\bea}{\begin{eqnarray}}
\newcommand{\eea}{\end{eqnarray}}

\newcommand{\p}{\partial}

\newcommand{\la}{\langle}
\newcommand{\ra}{\rangle}

\newcommand{\lp}{\left(}
\newcommand{\rp}{\right)}

\renewcommand{\vec}[1]{{\bf #1}}

\begin{document}
\title{Hot Carrier Transport and Photocurrent Response in Graphene}


\author{Justin C. W. Song$^{1,2}$}
\author{Mark S. Rudner$^3$}
\author{Charles M. Marcus$^3$}
\author{Leonid S. Levitov$^1$}
\affiliation{$^1$ Department of Physics, Massachusetts Institute of Technology, Cambridge, Massachusetts 02139, USA}
\affiliation{$^2$ School of Engineering and Applied Sciences, Harvard University, Cambridge, Massachusetts 02138, USA}
\affiliation{$^3$ Department of Physics, Harvard University, Cambridge, Massachusetts 02138, USA}





\begin{abstract}
{\bf ABSTRACT } Strong electron-electron interactions in graphene are expected to result in multiple-excitation generation by the absorption of a single photon. We show that the impact of carrier multiplication on photocurrent response is enhanced by very inefficient electron cooling, resulting in an abundance of hot carriers. The hot-carrier-mediated energy transport dominates the photoresponse and manifests itself in quantum efficiencies that can exceed unity, as well as in a characteristic dependence of the photocurrent on gate voltages. The pattern of multiple photocurrent sign changes 
as a function of gate voltage provides a fingerprint of hot-carrier-dominated transport and 
carrier multiplication.
\end{abstract} 
\maketitle

Graphene, a two-dimensional material with a gapless electronic spectrum, possesses a unique combination of optoelectronic characteristics \cite{bonaccorso}over a wide range of frequencies \cite{dawlaty, nair}.  Its high mobility enables high-speed photodetection\cite{xia2,mueller}. Combined with optical transparency and gate-tunable carrier density (field effect), this makes graphene an attractive material for photonic and optoelectronic applications\cite{mele,syzranov}. One particularly appealing aspect of photoresponse in graphene, anticipated by theory but yet to be confirmed experimentally, is the possibility of multiple-excitation generation by a single photon\cite{winzer}. This phenomenon, known as carrier multiplication (CM), is of fundamental importance for optoelectronics, since many optoelectronic devices can achieve much higher efficiencies if operated in the CM regime. Although CM has been reported in nano-particles such as colloidal PbSe and CdSe quantum dots\cite{klimov,califano,ellingson}, at present the interpretation of these measurements remains controversial \cite{bawendi}.

\begin{figure}
\includegraphics[scale=0.4]{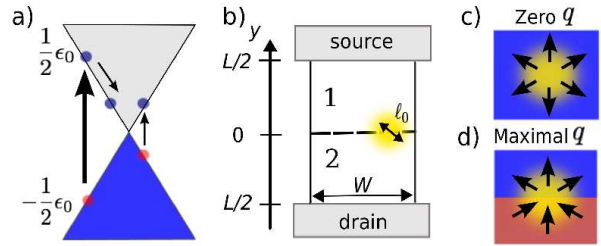}
\caption{(a) Carrier multiplication in graphene via impact ionization of secondary electron-hole pairs. (b) Schematic of an optoelectronic device with two separately gated regions 1 and 2, 
a laser excitation region positioned at the 1-2 interface, and a pair of contacts for collecting photocurrent. (c,d) Photoresponse in homogeneous (c) and inhomogeneous (d) systems. Arrows indicate thermoelectric current due to photogenerated hot carriers. Maximal quantum efficiency is achieved for opposite carrier polarities in regions 1 and 2, indicated by different colors (d).}
\label{fig1}
\vspace{-6mm}
\end{figure}

Recent theory suggests that CM can be readily achieved in graphene \cite{winzer}. Multiple-carrier production results from impact ionization and Auger-type processes induced by photoexcited carriers \cite{Rana07} (see Figure \ref{fig1}a). Fast carrier scattering, which dominates over electron-phonon scattering, is predicted by theory\cite{hwang-lifetime,kim} and confirmed by ultrafast dynamics observed in optical and terahertz pump-probe studies \cite{dawlaty2,george}. We distinguish the CM effect in graphene as `intrinsic' from the effects in avalanche photodiodes, which operate at relatively high reverse bias, sometimes just below breakdown. 

Here we set out to understand the relation between CM and photocurrent response in graphene. We argue that the contribution of CM processes, which occur locally in the photoexcitation region, is enhanced by energy transport throughout the entire system area. In particular, because of slow electron-lattice relaxation which serves as a bottleneck process for electron cooling\cite{macdonald}, photogenerated carriers are thermally decoupled from the crystal lattice over length scales which, even at room temperature, can be as large as $\xi\sim 7\,{\rm \mu m}$. Thermoelectric currents, arising in the presence of hot carriers, can lead to a dramatic enhancement of photoresponse.

As we shall see, these effects can have a direct impact on the  quantum efficiency of photoresponse, namely the number of photogenerated carriers transmitted through the contacts per absorbed photon, $q =N_{\rm el}/N_{\rm ph}$. This quantity is a cumulative characteristic of the measured photoresponse, which depends on various effects occurring throughout the system, including the CM processes in the excitation region, as well as charge and energy transport from this region to the contacts. These processes are characterized by very different time scales: fast generation of carriers by the CM process is followed by a much longer ``charge harvesting'' stage dominated by the drift of carriers from the excitation region to the contacts.

We emphasize that the hot-carrier regime arising under CM conditions is distinct from the effects of overall heating. In the latter case, since the electron heat capacity is very small compared to the lattice heat capacity, only a small fraction of the absorbed photon energy, equipartitioned between all degrees of freedom, would remain in the electron subsystem. As discussed in more detail below, this would result in a vanishingly small temperature change, and suppression of the hot-carrier effects. 

In contrast, slow electron-lattice relaxation triggers thermal imbalance of the electron and lattice subsystems, amplifying the CM effects. The electron-lattice relaxation slows down for temperatures below the Debye temperature. Under these conditions, the extreme inefficiency of cooling mediated by acoustic phonons allows the carriers to remain hot during their entire lifetime before reaching the contacts. Proliferation of hot carriers dramatically alters the nature of photoresponse.

We exhibit the essential physics of photoresponse by considering a double-gated device comprising two regions with gate-tunable carrier densities (see Figure \ref{fig1}b). As we will show, the quantum efficiency $q $ for this system has a simple dependence on the local electrical conductivities $\sigma_{1,2}$ and chemical potentials $\mu_{1,2}$ in the two regions:
\be\label{eq:Xi}
q = \frac{\alpha \epsilon_0}{(\sigma_1+\sigma_2)^2}\lp \sigma_2 \frac{\p\sigma_1}{\p\mu_1} -\sigma_1\frac{\p\sigma_2}{\p\mu_2} \rp
,\quad
\epsilon_0=hf
,
\ee
where the factor $\alpha$ ($0<\alpha<1$) describes the net fraction of the photon energy $\epsilon_0$ which is transferred to the electron system through photoexcitation 
and subsequent decay. Remarkably, under realistic conditions $q $ does not depend on device dimensions and temperature, and can take values as high as $q \gtrsim 2$ (see detailed estimate below).

We characterize the electron system by an electron temperature, $T_{\rm el}$, which in general is different from the lattice temperature $T_0$. The electron energy distribution is established via electron-electron scattering which occurs on a sub-picosecond time scale\cite{hwang-lifetime}. Since these times are shorter than the electron-phonon timescales, the electron-lattice relaxation can be described by a two-temperature model. Crucially, the processes due to optical phonons, which occur on relatively short times of several picoseconds\cite{dawlaty2}, become {\it quenched} when the photogenerated carrier energies drop well below the Debye energy, $\omega_{\rm D} \approx 200\, {\rm meV}$\cite{efetov}. For the carriers with lower energies, the dominant cooling process is mediated by acoustic phonons, giving a slow cooling rate. The cooling rate mediated by the absorption and emission of an acoustic phonon was studied by Bistritzer and MacDonald\cite{macdonald}. For cooling rate defined as $dT_{\rm el}/dt=-\gamma_1(T_{\rm el}-T_0)$, $T_{\rm el}\approx T_0$, they find
\be\label{eq:gamma}
\gamma_1=\frac{3V^2 \mu^3}{4\pi^2\hbar^3\rho v_F^4 k_{\rm B}T_{\rm el}}
\approx 0.87 \frac{\lp\mu\,[{\rm meV}]/100\rp^3}{T_{\rm el} \,[{\rm K}] /300 } {\rm  ns^{-1}} 
,
\ee
where $V=20\,{\rm eV}$, $\rho = 7.6 \times 10^{-8}\, {\rm g}/{\rm cm}^{2}$, and $v_F = 10^6\, {\rm m}/{\rm s}$ are the deformation potential, mass density, and Fermi velocity for graphene. For realistic values $\mu=100\, {\rm meV}$  and $T_{\rm el} \sim 300K$, we arrive at timescales as long as a few nanoseconds.

\begin{figure}
\includegraphics[scale=0.35]{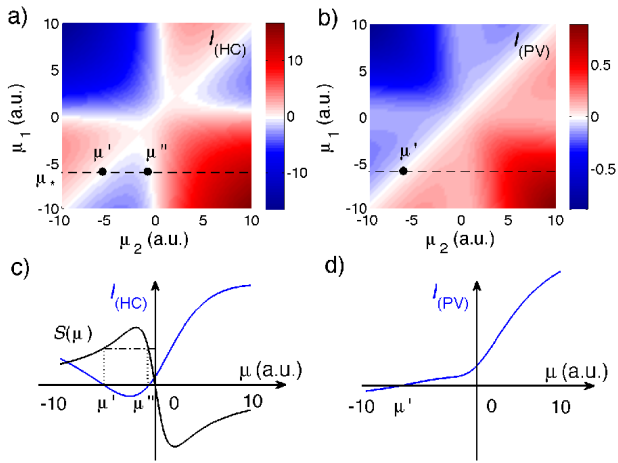}
\caption{
Photocurrent map as a function of chemical potentials $\mu_{1}$ and $\mu_{2}$ for the device shown in Figure \ref{fig1}b. Separately shown are the HC contribution (a) and the PV contribution (b), described by Eq.(\ref{eq:I=Sa-Sb}) and Eq.(\ref{pv2}), respectively. Note multiple polarity changes for the HC contribution as opposed to a single polarity change for the PV contribution. The scales of $I_{\rm (HC)}$ and $I_{\rm (PV)}$ have been calibrated to agree with the ratio calculated in Eq.(\ref{estimate}) with $L= 6\,{\rm \mu m}$.
Current slices along the dotted lines $\mu_1=\mu_*$ shown for the HC contribution (c) and the PV contribution (d). The Seebeck coefficient $s(\mu)$ is shown in (c).}
\label{polarity}
\vspace{-6mm}
\end{figure}

Slow cooling results in {\it thermal decoupling} of the electrons from the crystal lattice and energy transport mediated by hot carriers. As we shall see, the effects due to energy transport dominate over the conventional photovoltaic contribution to photoresponse. The hot-carrier mechanism and the photovoltaic mechanism of photoresponse have very  different experimental signatures. In particular, hot carriers manifest themselves in {\it multiple sign reversals of photoresponse} as carrier concentration is tuned by a monotonic sweep of gate voltage. Multiple sign changes do not occur in the absence of CM (see Fig.\ref{polarity}). Thus the pattern of photocurrent sign changes provides a fingerprint that can be used to experimentally identify the hot-carrier regime and the presence of strong CM.

We describe the electric current in the hot-carrier regime through the local current density
\be\label{eq:current}
\vec j=\sigma\vec E-e\eta n_{\rm x}(\vec r)\nabla U_g(\vec r)+\sigma s\nabla T_{\rm el}
.
\ee
The first two terms describe the conventional photovoltaic (PV) effect: primary photogenerated carriers are accelerated by the gate-induced electric field $-\nabla U_g$, and create a local photocurrent in the excitation region (here $n_{\rm x}$ is the steady state density of photoexcited carriers and $\eta$ is the mobility at energy $\epsilon\sim\frac12hf$). The redistribution of carriers associated with this local photocurrent sets up an electric field $\vec E=-\nabla(\phi-\mu/e)$ that drives current outside the excitation region, reaching the contacts. The last term in Eq.(\ref{eq:current}) describes the contribution of energy transferred to electrons via multiple-carrier generation, which takes the form of a thermoelectric current driven by the electron temperature gradient. 

The quantities $s$ and $\sigma$ in Eq.(\ref{eq:current}) are the Seebeck coefficient and electrical conductivity, which depend on local carrier density and sign. The temperature profile can be found using the energy flux 
\be\label{eq:energy_flux}
\vec W=\lp \phi-\frac{\mu}{e}\rp\vec j-\Pi\vec j-\kappa\nabla T_{\rm el}
,
\ee
where $\Pi=s T$ and $\kappa$ are the Peltier and the thermal conductivity coefficients. The values of $\sigma$, $s$ and $\kappa$ depend on the microscopic scattering mechanisms. In the practically interesting regime of disorder-dominated scattering, we have 
\be\label{eq:S_kappa}
s =\frac{\la (\epsilon-\mu)\tau\ra}{eT_{\rm el}\la \tau\ra}
,\quad
\kappa=\frac{\sigma}{e^2T_{\rm el}}\lp \frac{\la \epsilon^2\tau\ra}{\la\tau\ra}-\frac{\la\epsilon\tau\ra^2}{\la\tau\ra^2}\rp
,
\ee
where $\la ...\ra$ denotes averaging over the energy distribution of carriers, and $\tau(\epsilon)$ is the mean free scattering time for elastic collisions. 

We consider a simple model of photocurrent generation in graphene, based on Eqs.(\ref{eq:current}),(\ref{eq:energy_flux}), which accounts for the multiscale character of photoresponse: fast carrier kinetics within a micron-size excitation region set up a pattern of local electron temperature and electric fields that drive current throughout the entire device. As shown in Figure \ref{fig1}b, we consider a rectangular graphene sample of width $W$ and length $L$, with a step in carrier density at the interface between regions 1 and 2.
Photocurrent is collected through two contacts placed at $y=\pm L/2$.

In this model, using the continuity relation $\nabla\cdot\vec j=0$ and Eq.(\ref{eq:current}), we express the photocurrent as
\be\label{eq:I_general}
I=\int_0^W\!\!\int_{-L/2}^{L/2} \!\! \lp s(y)\nabla T_{\rm el} -\sigma^{-1}(y)e\eta n_{\rm x}\nabla U_{\rm g}\rp \frac{dy dx}{RW} 
\ee
where $R=\frac1{W}\int_{-L/2}^{L/2}\sigma^{-1}(y)dy$ is the total resistance, and the contacts are taken to be at equal potentials, $\int_{-L/2}^{L/2}E_ydy=0$. We focus on the term $s(y)\nabla T_{\rm el}$ in Eq.(\ref{eq:I_general}), and for the time being ignore the other term. The latter contribution will be analyzed below and shown to be small. Approximating the dependence $s(y)$ by a step that mimics the density profile, we express the hot-carrier (HC) contribution through the average increase of the electron temperature along the 1-2 interface
\be\label{eq:I=Sa-Sb}
I_{\rm (HC)}=
\frac{s_1-s_2}{R} \Delta T
,\quad
\Delta T=T_{{\rm el},\, y=0}^{\rm ave}-T_0
,
\ee
where $s_1$ and $s_2$ are the Seebeck coefficients in regions 1 and 2, $T_0$ is the lattice temperature, and the superscript `ave' stands for the value averaged over $0<x<W$, $y=0$. The spatial profile of $T_{\rm el}$ must be determined from the heat transport  equation 
\be\label{eq:heat_transport}
\nabla\cdot\vec W +\gamma C_{\rm el}(T_{\rm el}-T_0)=\alpha \epsilon_0 \dot N
,\quad
\epsilon_0=hf
,
\ee 
where $\gamma$ is the electron-lattice cooling rate (\ref{eq:gamma}), $C_{\rm el}$ is the electron specific heat, and $\dot{N}$ is  the photon flux absorbed in the laser spot. Since typical spot sizes $ \sim 0.5 - 1  \,\mu {\rm m}$ \cite{xu} are smaller than other relevant length scales, such as the system size and cooling length (see below), the absorbed photon flux can be approximated as a delta-function source of hot electrons.

Next, we analyze the HC photocurrent dependence on gate voltages of regions 1 and 2. Two simple cases are illustrated in Figures.~\ref{fig1}c,d: the net HC current vanishes for a spatially uniform carrier density, and is maximized for a p-n interface. The full dependence on the chemical potentials $\mu_{1,2}$, illustrated in Figure \ref{polarity}a, shows stronger photoresponse for $\mu_1$ and $\mu_2$ of opposite sign. This is in agreement with the `gate-activated photoresponse' observed in Ref.\cite{frank} in the presence of a gate-induced p-n junction.

We model the dependence on the chemical potentials $\mu_{1,2}$, given by the factor $\frac{s_1-s_2}{R}$ in Eq.(\ref{eq:I=Sa-Sb}), using $R=\frac{L}{2W}\frac{\sigma_1+\sigma_2}{\sigma_1 \sigma_2}$ and the Mott formula\cite{hwang_thermopower} for the Seebeck coefficient obtained from the non-interacting model (\ref{eq:S_kappa}),
\be\label{eq:sigma_S}
s(\mu) =  - \frac{\pi^2 k_{\rm B}^2T}{3e }\frac1{\sigma} \frac{d\sigma}{d \mu}
,\quad
\sigma(\mu) = \sigma_{\rm min}\lp 1+\frac{\mu^2}{\Delta^2}\rp
,
\ee
where $k_{\rm B}T\ll {\rm max}(\mu,\Delta)$. Here $\sigma(\mu)$ describes a linear dependence of conductivity on carrier concentration away from the Dirac point, with parameters $\sigma_{\rm min}$, the minimum conductivity, and $\Delta$, the width of the neutrality region. The dependence $s(\mu)$, Eq.(\ref{eq:sigma_S}), is in good agreement with the measurements of thermopower in graphene\cite{zuev}.

The dependence of photocurrent on $\mu_1$ and $\mu_2$ has a number of interesting features. Because of the dependence on $s(\mu_1)-s(\mu_2)$, the HC current (\ref{eq:I=Sa-Sb}) vanishes on the diagonal $\mu_1=\mu_2$ and, in addition, on two hyperbolae $\mu_1\mu_2=\Delta^2$ that cross the diagonal. We shall refer to the latter as `anomalous' polarity reversal. As illustrated in Figure \ref{polarity}c, this behavior can be traced to the non-monotonic character of the dependence $s(\mu)$. In particular, for any nonzero value of $\mu_1$ excluding extrema of $s(\mu)$, the dependence on $\mu_2$ features two polarity changes. This is illustrated by the slice $\mu_1=\mu_*$ in Figure \ref{polarity}c. At the nodes $\mu_2=\mu',\mu''$, the Seebeck coefficient satisfies $s(\mu_*)=s(\mu')=s(\mu'')$, as shown by the horizontal dashed line in Figure \ref{polarity}c. Hence the photocurrent has opposite polarity inside and outside the interval $\mu'< \mu_2 < \mu''$. Using a numerical evaluation of the integral in Eq.(\ref{eq:I=Sa-Sb}), we checked that multiple polarity reversal, `proper' and `anomalous', as well as other qualitative features are insensitive to the photoexcitation spot size, surviving even for spatially uniform photoexcitation. 

A model similar to Eq.(\ref{eq:I=Sa-Sb}) was used in Ref.\cite{xu} to describe a laser-induced photo-thermoelectric effect observed in a heterogeneous system, a monolayer-bilayer interface. The measured photocurrent sign was consistent with Eq.(\ref{eq:I=Sa-Sb}) but not with the PV effect (see below).

The multiple polarity reversal in the dependence on $\mu_1$ and $\mu_2$ is unique for the HC mechanism. It is instructive to make a comparison with the photocurrent response in the conventional photovoltaic (PV) regime where the primary photogenerated pair is the main contributor to photoresponse. This contribution is described by the second term of Eq.(\ref{eq:I_general}), giving
\be\label{pv2_general}
I_{\rm (PV)}=-\frac1{RW}\int\!\!\int \sigma^{-1}(\vec r) e\eta n_{\rm x}(\vec r)  \nabla U_g(\vec r)\, dx dy
.
\ee
The integration simplifies when the size of the photoexcitation spot is larger than the depletion length, $l_0\gg w_d$. Setting $eU_g(\vec r)=\mu(\vec r)$, using the model dependence $\sigma(\mu)$ from Eq.(\ref{eq:sigma_S}), and replacing the integration over $y$ by integration over $\mu$, gives 
\be\label{pv2}
 I_{\rm (PV)}= \frac{\eta\Delta}{\sigma_{\rm min}R}\lp \tan^{-1}\frac{\mu_1}{\Delta} -\tan^{-1}\frac{\mu_2}{\Delta} \rp  n_{\rm x}^{\rm ave} (y=0)
.
\ee
Result (\ref{pv2}) also shows photoresponse maximized in the presence of a p-n junction, i.e. for $\mu_1$ and $\mu_2$ of opposite sign. Thus the `gate-activated photoresponse' of Ref.\cite{frank} cannot be used to distinguish the HC and PV contributions to photocurrent. 

The difference between the two contributions to photocurrent is most striking when $\mu_1$ and $\mu_2$ have equal signs [see Figure \ref{polarity}a,b]. Since the polarity of the PV current is determined solely by the sign of field gradient $\nabla U_g$, there is only one sign reversal occurring at $\mu_1=\mu_2$. In contrast, the nonmonotonic character of $s(\mu)$ produces multiple polarity reversal of the HC contribution. Thus the polarity of photocurrent as a function of gate potentials offers a direct way to differentiate between the two mechanisms.

To estimate the magnitude of the photocurrent, Eq.~(\ref{eq:I=Sa-Sb}), we need to obtain the steady state profile of $T_{\rm el}$ from Eq.(\ref{eq:heat_transport}). Since $\vec j$ has zero divergence, we can write 
\be
\nabla\cdot\vec W=-\sigma^{-1}|\vec j|^2-\vec j\cdot\nabla\Pi-\nabla\cdot(\kappa\nabla T).
\ee
The first term, which is quadratic in $\vec j$, can be ignored. The second term describes the Peltier cooling effect due to photocurrent passing through the 1-2 interface. Incorporating it in Eq.(\ref{eq:heat_transport}) gives
\be\label{eq:T_profile}
-\nabla\cdot(\kappa\nabla T_{\rm el}) +\gamma C_{\rm el} (T_{\rm el}-T_0) =\alpha \epsilon_0 \dot N+\vec j\cdot\nabla\Pi
.
\ee
Since the spatial extent of the Peltier term $\vec j\cdot\nabla\Pi$ is of order of the the p-n junction width, which is $\lesssim 0.1 \, \mu {\rm m}$ in the state-of-the-art devices, it can be well approximated as a delta function source localized at the 1-2 interface. 

Eq.(\ref{eq:T_profile}) can be conveniently analyzed using quantities averaged over the device width $0<x<W$, $T^{\rm ave}_{\rm el}(y)=\frac1{W}\int_0^WT_{\rm el}(x,y)dx$, $\dot{N}^{\rm ave}(y)=\frac1{W}\int_0^W\dot{N}(x,y)dx$, and transforming Eq.(\ref{eq:T_profile}) to a one-dimensional equation. For simplicity, we will consider the case when the laser spot is positioned on the 1-2 interface. Treating it as a delta function, we solve Eq.(\ref{eq:T_profile}) by piecing together solutions of the homogeneous equation satisfying zero boundary condition at the contacts $y=\pm L/2$, and performing matching of the boundary values at the 1-2 interface.

Estimating the cooling length $\xi=\sqrt{\kappa/\gamma C_{\rm el}} $ we find a large value $\xi\approx 7\,\mu{\rm m}$ which exceeds $L/2$ for typical device dimensions. Eq.(\ref{eq:T_profile}) can thus be solved by approximating $\gamma\approx 0$, yielding temperature profile $\delta T^{\rm ave}_{\rm el}(y)=(1-2|y|/L)\Delta T$ with the average temperature increase at the 1-2 interface
\be\label{eq:deltaT}
\Delta T = \frac{\alpha \epsilon_0 l_0 \dot{N}^{\rm ave}_{y=0}}{\frac2{L}(\kappa_1+\kappa_2)+\frac{T_0}{RW}(s_1-s_2)^2}
.
\ee
Since $R\propto L$, $\Delta T$ grows linearly with system size, saturating when $L/2$ exceeds the cooling length $\xi$. For not too high temperatures, $k_{\rm B}T\lesssim {\rm max}(\mu_{1,2},\Delta)$, the second term in the denominator is smaller than the first term and can thus be ignored. Combining this result with Eq.(\ref{eq:I=Sa-Sb}), we evaluate the quantum efficiency as $q =I_{\rm (HC)}/(e l_0 W \dot{N}^{\rm ave}_{y=0})$. Using the Wiedemann-Franz (WF) relation $e^2\kappa =\frac{\pi^2}3k_{\rm B}^2T\sigma$, and the Mott formula, Eq.(\ref{eq:sigma_S}), we arrive at Eq.(\ref{eq:Xi}). The result (\ref{eq:Xi}) describes the realistic situation of large cooling length, $\xi\gtrsim L$. 

For a general system size $L$, the solution can be obtained as $\delta T(y<0)=A_1\sinh((y+\frac12L)/\xi_1)$, $\delta T(y>0)=A_2\sinh((\frac12 L-y)/\xi_2)$. After matching the boundary values and derivatives at $y=0$, we obtain
\be
\Delta T = \frac{\alpha \epsilon_0 l_0 \dot{N}^{\rm ave}_{y=0}}{ \frac{\kappa_1}{\xi_1}{\rm coth}\frac{L}{2\xi_1}+\frac{\kappa_2}{\xi_2}{\rm coth}\frac{L}{2\xi_2}+\frac{T_0}{RW}(s_1-s_2)^2}
.
\ee
This result agrees with Eq.(\ref{eq:deltaT}) for small system size $L\ll\xi_{1,2}$. At large $L$, it describes saturation to the value $\Delta T = \alpha \epsilon_0 l_0 \dot{N}^{\rm ave}_{y=0}/(\kappa_1/\xi_1+\kappa_2/\xi_2)$.

For an estimate of the numerical value of $q $, we use the factor $\alpha$ calculated in Ref.\cite{winzer} which predicts $\alpha \epsilon_0 = Mk_{\rm B}T_0$ with $M=4.3$, $T_0=300\,{\rm K}$, thus giving $\alpha\approx 0.07$. Taking the chemical potentials at the minimum and maximum of $s(\mu)$ described by the model (\ref{eq:sigma_S}), $\mu_{1,2}=\pm\Delta$, we obtain $q =M k_{\rm B}T_0/2\Delta$. 
Taking a typical value for the neutrality region width for graphene on BN substrate, $\Delta\lesssim 300\,{\rm K}$, we find $q \gtrsim 2$. Thus, high quantum efficiencies are feasible for realistic system sizes of up to $5-10\,{\rm \mu m}$.

We now proceed to estimate the relative strength of the HC and PV contributions to photoresponse. Using the WF relation and the Mott formula, Eq.(\ref{eq:sigma_S}), we find
\be\label{eq:I_pt/I_pv}
\frac{I_{\rm (HC)}}{I_{\rm (PV)}}\approx \frac{\sigma_{\rm min}s l_0 L \alpha\epsilon_0\dot N^{\rm ave}}{2\Delta \kappa\eta n_{\rm x}^{\rm ave}}
=\frac{e\alpha\epsilon_0 l_0 L}{4\eta\tau_0\Delta }\times \frac{\sigma_{\rm min}}{\sigma^2}\frac{d\sigma}{d\mu}
,
\ee
where we estimated the photoexcited carrier density as $n_{\rm x}=2\tau_0\dot N$ with $\tau_0$ the carrier lifetime. Near the Dirac point, estimating $\frac{\sigma_{\rm min}}{\sigma^2}\frac{d\sigma}{d\mu}\sim 1/\Delta$, we obtain
\be
\label{estimate}
\frac{I_{\rm (HC)}}{I_{\rm (PV)}}\approx\frac{e\alpha\epsilon_0 l_0 L}{4\eta\tau_0\Delta^2 }\approx   2.6 L\,[{\rm \mu m}] \sim 15-25
\ee
for $L\sim 5-10\,{\rm \mu m}$, where we used parameter values: $\eta = 10^4 \,\textrm{cm}^2 V^{-1}s^{-1}$, $\tau_0= 1\,\textrm{ps}$\cite{george}, the neutrality region width $\Delta\approx 100\,{\rm meV}$ estimated for graphene on SiO substrate\cite{jmartin}, $\alpha \approx 0.07$ \cite{winzer}, the photon energy $\epsilon_0 \sim 1.5\,\textrm{eV}$, and the laser spot size $l_0 \approx 1\,\mu{\rm m}$ \cite{xu}.

We therefore conclude that, due to very inefficient electron-lattice cooling (Eq.(\ref{eq:gamma})) and efficient CM process, an abundance of hot carriers leads to a dominant HC contribution to the photoresponse.  Furthermore, because the ratio $I_{\rm (HC)}/I_{\rm (PV)}$ scales inversely with the square of the neutrality region width $\Delta^2$, we expect hot-carrier-related phenomena such as high quantum efficiency to become more pronounced for high mobility samples, e.g. graphene on a boron nitride substrate. 

To better understand the relation between the result (\ref{estimate}) and CM, it is instructive to consider the situation when energy is equally partitioned between all degrees of freedom, electrons and lattice (which would be the case for very fast electron-lattice relaxation). Crucially, the large difference between the electron and phonon specific heat values makes the lattice act as a nearly ideal heat sink. In this case,
the value $\alpha\approx 0.07$ used above, describing the fraction of photon energy remaining in the electron subsystem, would be replaced by the heat capacity ratio $C_{\rm el}/C _{\rm ph}$, which is very small. In the idealized case of a sharp Dirac point and undoped system, we have $C_{\rm el}/C _{\rm ph} =v_{ph}^2/v_{\rm F}^2 \approx 10^{-4}$. Although $C_{\rm el}$ is somewhat enhanced at finite doping, under realistic conditions it remains quite small. We estimate $C_{\rm el}/C _{\rm ph} \approx (v_{ph}^2/v_{\rm F}^2) \times (\mu/k_BT)$, where $k_BT \ll\mu$. Taking $\mu=100\, {\rm meV}$, for temperatures $10\,{\rm K}<T<300\,{\rm K}$ we obtain $C_{\rm el}/C _{\rm ph} \approx 10^{-2} -10^{-4}$. This would reduce our estimate of the ratio  $I_{\rm (HC)}/I_{\rm (PV)}$ by a large factor $10 - 10^3$, strongly suppressing $I_{\rm (HC)}$. Hence, a dominant $I_{\rm (HC)}$ signals that the electronic system and lattice are out of equilibrium viz. the presence of CM.

In summary, hot-carrier transport in the presence of photoexcitation leads to a novel type of photoresponse dominated by photo-thermoelectric effects. Multiple sign reversals of photocurrent over a monotonic sweep of gate voltage is a hallmark of this regime. The pattern of sign changes provides a clear fingerprint for experimentally identifying efficient carrier multiplication. Better understanding of nonequilibrium transport physics in graphene, and in particular of different pathways for cooling, is needed to develop a detailed picture of this new regime. This will open up broad vistas for both the exploration of novel opto-electronic properties as well as the design of more efficient graphene photodetectors.

\section*{Acknowledgement}
We thank  N. Gabor, Nan Gu, P. Jarillo-Herrero, F. Koppens, M. Lemme,
and R. Nandkishore for useful discussions, and acknowledge
support from the NSS program, Singapore (JS) and Naval Research Grant N00014-09-1-0724 (LL).

\vspace{-5mm}
\section*{REFERENCES AND NOTES}

\end{document}